\documentclass[12pt,onecolumn,showpacs,preprintnumbers,amsmath,amssymb,fleqn]{revtex4}
\usepackage{graphics}
\normalsize

\begin{document}

\title{Two atoms in dissipative cavities in dispersive limit: entanglement
sudden death and long-lived entanglement}

\author{Jian-Song Zhang$^{1}$}
\email{jszhang1981@zju.edu.cn}
\author{Ai-Xi Chen$^{1}$}
\author{M. Abdel-Aty$^{2}$}
\affiliation{$^{1}$Department of Applied Physics, East China Jiaotong University,
Nanchang 330013, People's Republic of China \\
$^{2}$Mathematics Department, Faculty of Science, Sohag University, 82524
Sohag, Egypt}

\begin{abstract}
We investigate the entanglement dynamics and coherence of two two-level
atoms interacting with two coherent fields of two spatially separated and
dissipative cavities. It is in particular shown that entanglement sudden
death is obtained within a short interacting time. However, after a long
interaction time a long-lived entanglement is shown, that is, the initial
entanglement of two atoms could be partially preserved. In addition, the
coherence of the two atoms will not be lost during the evolution.

\pacs{03.67.Mn; 03.65.Ud}

Keywords: Cavity QED; Dissipation; Entanglement dynamics; Coherence
\end{abstract}
\maketitle

\section{Introduction}

Quantum entanglement is at the heart of quantum information processing and
quantum computation \cite{Nielsen2000,Raimond2001,Yang2005,Kim2002}. In recent years,
many efforts have been devoted to the study of the evolution of joint
systems formed by two qubits \cite{Yu2004,Yu2009,Jakobczyk2004,Ficek2006,Yonac,
Yu2007,Ikram2008,Bellomo2007,Bellomo2008,Rau2008,Jamroz2006}. In particular,
Yu and Eberly \cite{Yu2004,Yu2009} have found out that the single-qubit dynamics
and the global dynamics of an entangled two-qubit system subjected to
independent environments are rather different. For a single-qubit system
subjected to an environment, the local coherence decays asymptotically while
the entanglement of an entangled two-qubit system may disappear for a finite
time during the dynamics evolution. The nonsmooth finite-time disappearance
of entanglement is called \textquotedblleft entanglement sudden
death\textquotedblright\ (ESD). More recently, the ESD phenomenon has been
observed in the laboratory by two groups \cite{Almeida2007,Laurat2007}.

One the other hand, the long-lived entanglement in cavity QED or
solid state systems was investigated by several authors
\cite{Yu2006,Xu2005,Aty2006,Dajka2007,Aty2008}. In Ref.
\cite{Yu2006}, the authors have discussed the effects of the
classical noise on entanglement dynamics and pointed out that there
is long-lived entanglement in the presence of global dephasing
noise. The influence of white noise on the entanglement dynamics of
two atoms within two cavities has been investigated by Xu and Li
\cite{Xu2005}. They found out that if only one atom is driven by the
white-noise field, then there is long-lived entanglement. The
entanglement dynamics in two effective two-level trapped ions
interacting with a laser field has been studied in \cite{Aty2006}.
It has been shown by Dajka, Mierzejewski, and Luczka that there is
long-lived entanglement of some composite systems even if coupled
with a thermal bath \cite{Dajka2007}. It has also been pointed out
that phase-damped cavity can lead to long-lived entanglement in a
quantum system consisting of a single-Cooper-pair box irradiated by
a quantized field \cite{Aty2008}.

In the present paper, we investigate the entanglement dynamics and
coherence of a quantum system formed by two two-level atoms within
two spatially separated and dissipative cavities in the dispersive
limit by the employing concurrence and linear entropy, respectively.
The two atoms are initially entangled and the cavities are initially
prepared in coherent states. We show that the ESD phenomenon appears
in the present system and the entanglement of two atoms decreases
with the time in the short-term. However, the long-term behavior is
very different since a long survival of the entanglement is shown in
the system. In other words, there is long-lived entanglement
(stationary state entanglement) in the presence of the dissipation
of cavities, implying that the initial entanglement of two atoms
could be partially preserved even they are put into dissipative
cavities. We find the long-lived entanglement depends on the initial
state of the two atoms and the ratio of the decay rate of the
cavities to the atom-field coupling constant. Finally, we discuss
the coherence of the two atoms using the linear entropy. Our results
show that the linear entropy of each atom at any time is equal to
the initial linear entropy, that is, the coherence of each atom is
preserved. We find that the initial coherence (entanglement) of the
two atoms could be preserved (partially preserved) in the dispersive
limit even if they are put into two dissipative cavities.

The present paper is organized as follows. In Section II, we obtain an
explicit analytical solution of one atom interacting with a dissipative
cavity in the dispersive limit. In section III, we consider a quantum system
consisting of two atoms within two spatially separated cavities. In Section
IV, the entanglement dynamics and coherence of the two two-level atoms are
investigated by employing the concurrence and linear entropy, respectively.
Finally, we summarize our results in Section V.

\section{One atom in a dissipative cavity}
We first consider a quantum system consisting of a two-level atom
interacting with a single-mode cavity. Under the electric dipole and
rotating wave approximation, the Hamiltonian of the present system is $%
(\hbar =1)$ \cite{Scully1997,Walls1995}
\begin{eqnarray}
H=\omega a^{\dag }a+\frac{\omega _{0}}{2}\sigma _{z}+g(a^{\dag }\sigma
_{-}+a\sigma _{+}),
\end{eqnarray}
where $g$ is the atom-field coupling constant, $\sigma _{z}=|e\rangle
\langle e|-|g\rangle \langle g|$ and $\sigma _{\pm }$ are the atomic spin
flip operators characterizing the effective two-level atom with frequency $%
\omega _{0}$. The symbols $|e\rangle $ and $|g\rangle $ refer to the excited
and ground states for the two-level atom. Here, $a^{\dag }$ and $a$ are the
creation and annihilation operators of the field with frequency $\omega $,
respectively. The dispersive limit is obtained when the condition $|\Delta
|=|\omega _{0}-\omega |\gg \sqrt{n+1}g$ is satisfied for any relevant $n$.
Then, the interaction Hamiltonian $g(a^{\dag }\sigma _{-}+a\sigma _{+})$ can
be regarded as a small perturbation and the effective Hamiltonian of the
model can be recast as \cite{Meystre1992}
\begin{eqnarray}
H_{e}=\omega a^{\dag }a+\frac{\omega _{0}}{2}\sigma _{z}+\Omega \lbrack
(a^{\dag }a+1)|e\rangle \langle e|-a^{\dag }a|g\rangle \langle g|],
\end{eqnarray}
with $\Omega =g^{2}/\Delta $. In the interaction picture, the interaction
Hamiltonian is
\begin{eqnarray}
V=\Omega \lbrack (a^{\dag }a+1)|e\rangle \langle e|-a^{\dag }a|g\rangle
\langle g|].
\end{eqnarray}
We assume the two-level atom interacting with a coherent field in a
dissipative environment. This interaction causes the losses in the cavity
which is presented by the superoperator $\mathcal{D}=k(2a\cdot a^{\dag
}-a^{\dag }a\cdot -\cdot a^{\dag }a)$, where $k$ is the decay constant. For
the sake of simplicity, we confine our consideration in the case of zero
temperature cavity. Then, the master equation that governs the dynamics of
the system can be written as follows
\begin{eqnarray}
\frac{d\rho ^{\prime }}{dt} &=&-i[V,\rho ^{\prime }]+\mathcal{D}\rho
^{\prime }  \nonumber \\
&=&-i[V,\rho ^{\prime }]+k(2a\rho ^{\prime }a^{\dag}-a^{\dag }a\rho ^{\prime
}-\rho ^{\prime }a^{\dag}a)\nonumber\\
&=&-i[V,\rho ^{\prime }]+k(2\mathcal{M}\rho ^{\prime }-\mathcal{R}\rho ^{\prime
}-\mathcal{L}\rho ^{\prime }),  \label{master}
\end{eqnarray}%
where $\rho ^{\prime }$ is the density matrix of the atom-field system
and $\mathcal{M}$, $\mathcal{R}$, and $\mathcal{L}$ are defined by
\begin{eqnarray}
\mathcal{M}\rho ^{\prime }=(a\cdot a^{\dag })\rho ^{\prime }=a\rho ^{\prime } a^{\dag },
\mathcal{R}\rho ^{\prime }=(a^{\dag }a\cdot)\rho ^{\prime }=a^{\dag }a\rho ^{\prime },
\mathcal{L}\rho ^{\prime }=(\cdot a^{\dag }a)\rho ^{\prime }=\rho ^{\prime }
a^{\dag }a.
\end{eqnarray}
Here the superoperators $a\cdot ,\cdot a,a^{\dag }\cdot $, and $\cdot a^{\dag
}$ represent the action of creation and annihilation operators on an
operator
\begin{eqnarray}
(a\cdot )o=ao,(\cdot a)o=oa,\quad (a^{\dag }\cdot )o=a^{\dag }o,(\cdot
a^{\dag })o=oa^{\dag }.
\end{eqnarray}

We can express the density matrix in the following form
\begin{eqnarray}
\rho ^{\prime }(t) &=&\rho _{ee}^{\prime }(t)\otimes |e\rangle \langle
e|+\rho _{gg}^{\prime }(t)\otimes |g\rangle \langle g|+\rho _{eg}^{\prime
}(t)\otimes |e\rangle \langle g|  \nonumber \\
&&+\rho _{ge}^{\prime }(t)\otimes |g\rangle \langle e|,  \label{e1}
\end{eqnarray}%
where $\rho _{ij}^{\prime }$'s are defined as $\rho _{ij}^{\prime }=\langle
i|\rho ^{\prime }|j\rangle $, $\rho _{ij}^{\prime }=\rho _{ji}^{^{\prime
}\dag }$, $i,j=e,g$. A straightforward calculation shows that
\begin{eqnarray}
\frac{d\rho ^{\prime }_{ee}(t)}{dt} &=&\{-i\Omega (\mathcal{R}-\mathcal{L})+k(2%
\mathcal{M}-\mathcal{R}-\mathcal{L})\}\rho _{ee}^{\prime }(t)=\mathcal{L}%
_{ee}\rho _{ee}^{\prime }(t),  \nonumber \\
\frac{d\rho ^{\prime }_{gg}(t)}{dt}  &=&\{i\Omega (\mathcal{R}-\mathcal{L})+k(2%
\mathcal{M}-\mathcal{R}-\mathcal{L})\}\rho _{gg}^{\prime }(t)=\mathcal{L}%
_{gg}\rho _{gg}^{\prime }(t),  \nonumber \\
\frac{d\rho ^{\prime }_{eg}(t)}{dt}  &=&\{-i\Omega (\mathcal{R}+\mathcal{L}+1)+k(2%
\mathcal{M}-\mathcal{R}-\mathcal{L})\}\rho _{eg}^{\prime }(t)  \nonumber \\
&=&\mathcal{L}_{eg}\rho _{eg}^{\prime }(t),  \nonumber \\
\rho _{ge}^{\prime }(t) &=&\rho _{eg}^{^{\prime }\dag }(t).
\end{eqnarray}%

It is easy to check that the superoperators $\mathcal{M}$, $\mathcal{R}$ and
$\mathcal{L}$ satisfy the relations
\begin{eqnarray}
\lbrack \mathcal{R},\mathcal{M}]=[\mathcal{L},\mathcal{M}]=-\mathcal{M},[%
\mathcal{R},\mathcal{L}]=0.
\end{eqnarray}
It is worth noting that $[\mathcal{R}+\mathcal{L},\mathcal{M}]=-2\mathcal{M}$%
, the superoperators $\mathcal{R}+\mathcal{L}$ and $\mathcal{M}$ form a
shift operator algebra. Thus we have the expansion of the exponential of a
linear combination of $\mathcal{R}+\mathcal{L}$ and $\mathcal{M}$
\begin{eqnarray}
e^{\mathcal{L}_{ee}t} &=&e^{(e^{2kt}-1)\mathcal{M}}e^{-(i\Omega +k)t\mathcal{%
R}}e^{(i\Omega -k)t\mathcal{L}},  \nonumber \\
e^{\mathcal{L}_{gg}t} &=&e^{(e^{2kt}-1)\mathcal{M}}e^{(i\Omega -k)t\mathcal{R%
}}e^{(-i\Omega +k)t\mathcal{L}}, \\
e^{\mathcal{L}_{eg}t} &=&e^{-i\Omega t}e^{[e^{2(i\Omega +k)t}-1]\mathcal{M}%
/(i\Omega +k)}e^{-(i\Omega +k)t\mathcal{R}}e^{-(i\Omega +k)t\mathcal{L}}.
\nonumber  \label{m2}
\end{eqnarray}

Suppose the atom is initially prepared in the state $\left(
\begin{array}{cc}
\xi _{a} & \xi _{c} \\
\xi _{c}^{\ast } & \xi _{b}%
\end{array}%
\right) $ and the field is initially prepared in the coherent state $|\alpha
\rangle =e^{-|\alpha |^{2}/2}\sum_{n=0}^{\infty }\frac{\alpha ^{n}}{\sqrt{n!}%
}|n\rangle $ with $\alpha $ being a complex number. Here, $|n\rangle $ is
the Fock state with $a^{\dag }a|n\rangle =n|n\rangle $. Therefor the initial
state of the atom-cavity system is
\begin{eqnarray}
\rho ^{\prime }(0) &=&\left(
\begin{array}{cc}
\xi _{a} & \xi _{c} \\
\xi _{c}^{\ast } & \xi _{b}%
\end{array}%
\right) \otimes |\alpha \rangle \langle \alpha |  \nonumber \\
&=&(\xi _{a}|e\rangle \langle e|+\xi _{b}|g\rangle \langle g|+\xi
_{c}|e\rangle \langle g|  \nonumber \\
&&\qquad +\xi _{c}^{\ast }|g\rangle \langle e|)\otimes |\alpha \rangle
\langle \alpha |.  \label{e2}
\end{eqnarray}%
Comparing Eq.(\ref{e1}) with Eq.(\ref{e2}), one get
\begin{eqnarray}
\rho _{ee}^{\prime }(0) &=&\xi _{a}|\alpha \rangle \langle \alpha |,\rho
_{gg}^{\prime }(0)=\xi _{b}|\alpha \rangle \langle \alpha |,  \nonumber \\
\rho _{eg}^{\prime }(0) &=&\xi _{c}|\alpha \rangle \langle \alpha |,\rho
_{ge}^{\prime }(0)=\rho _{eg}^{^{\prime }\dag }(0).  \label{e3}
\end{eqnarray}%
Combing Eqs.(10), (\ref{e3}) and the relation $\rho _{ij}^{\prime }(t)=e^{%
\mathcal{L}_{ij}t}\rho _{ij}^{\prime }(0)$, we find that the matrix elements
$\rho _{ij}^{\prime }(t)$ at time t is given by
\begin{eqnarray}
\rho _{ee}^{\prime }(t) &=&\xi _{a}|\alpha _{+}(t)\rangle \langle \alpha
_{+}(t)|,  \nonumber \\
\rho _{gg}^{\prime }(t) &=&\xi _{b}|\alpha _{-}(t)\rangle \langle \alpha
_{-}(t)|,  \nonumber \\
\rho _{eg}^{\prime }(t) &=&\xi _{c}f(t)|\alpha _{+}(t)\rangle \langle \alpha
_{-}(t)|,\rho _{ge}^{\prime }(t)=\rho _{eg}^{^{\prime }\dag }(t)  \nonumber
\\
|\alpha _{\pm }(t)\rangle  &=&|\alpha e^{-(k\pm i\Omega )t}\rangle ,
\nonumber \\
f(t) &=&\exp {\{-i\Omega t+|\alpha |^{2}(e^{-2kt}-1)\}}  \nonumber \\
&&\times \exp {\{\frac{|\alpha |^{2}k}{k+i\Omega }[1-e^{-2(k+i\Omega )t}]\}}.
\end{eqnarray}%
The reduced density matrix of the atom is obtained by tracing out the
variables of the field from the above density matrix
\begin{eqnarray}
\rho _{atom}^{\prime }(t) &=&\xi _{a}|e\rangle \langle e|+\xi _{b}|g\rangle
\langle g|  \nonumber \\
&&+[\xi _{c}f(t)\lambda (t)|e\rangle \langle g|+h.c],  \nonumber \\
\lambda (t) &=&\langle \alpha _{-}(t)|\alpha _{+}(t)\rangle,   \label{s1}
\end{eqnarray}%
where $h.c$ denotes the Hermitian conjugate.

\section{Two atoms within two spatially separated cavities}

In this section, we consider a quantum system consisting of two
noninteracting atoms each locally interacts with its own coherent field of
a dissipative cavity. The two atoms are initially prepared in entangled states
and the two fields are prepared in coherent states $|\alpha _{1}\rangle $
and $|\alpha _{2}\rangle $. For the sake of simplicity, we assume $\alpha
_{1}=\alpha _{2}$, the decay rates of the two cavities are equal, and the
atom-field coupling constants are the same. The schematic picture of the
present model is presented in Fig.1.

\subsection{Reduced density matrix of two atoms}
Using the method introduced by Bellomo, France, and Compagno \cite{Bellomo2008},
 we can obtain the reduced matrix density of two atoms
conveniently. In the basis $|1\rangle =|ee\rangle ,|2\rangle =|eg\rangle
,|3\rangle =|ge\rangle ,|4\rangle =|gg\rangle ,$ and using Eq. (\ref{s1}),
the reduced density matrix $\rho (t)$ for the two-atom system is
\begin{eqnarray}
\rho _{ii}(t) &=&\rho _{ii}(0),  \nonumber \\
\rho _{12}(t) &=&f(t)\lambda (t)\rho _{12}(0),  \nonumber \\
\rho _{13}(t) &=&f(t)\lambda (t)\rho _{13}(0),  \nonumber \\
\rho _{14}(t) &=&[f(t)\lambda (t)]^{2}\rho _{14}(0),  \nonumber \\
\rho _{23}(t) &=&|f(t)\lambda (t)|^{2}\rho _{23}(0),  \nonumber \\
\rho _{24}(t) &=&f(t)\lambda (t)\rho _{24}(0),  \nonumber \\
\rho _{34}(t) &=&f(t)\lambda (t)\rho _{34}(0),
\end{eqnarray}%
with $\rho _{ij}(t)=\rho _{ji}^{\ast }(t)$ and $i=1,2,3,4$. We would like to
point out that the above procedure allows us to obtain the reduced density
matrix of the two-atom system for any initial state of the two atoms.

\subsection{Extended Werner-like states}

We assume the initial states of the two-atom system are the extended
Werner-like (EWL) states \cite{Bellomo2008} defined by
\begin{eqnarray}
\rho _{\Phi } &=&p|\Phi \rangle \langle \Phi |+\frac{1-p}{4}I,  \nonumber \\
\rho _{\Psi } &=&p|\Psi \rangle \langle \Psi |+\frac{1-p}{4}I,  \nonumber \\
|\Phi \rangle  &=&\mu |eg\rangle +\nu |ge\rangle ,  \nonumber \\
|\Psi \rangle  &=&\mu |ee\rangle +\nu |gg\rangle ,
\end{eqnarray}%
where $p$ is a real number which indicates the purity of initial states, $I$
is a $4\times 4$ identity matrix, $\mu $ and $\nu $ are complex numbers with
$|\mu |^{2}+|\nu |^{2}=1$. The parameter $p$ is 1 for pure sates and 0 for
completely mixed states. It is worth noting that the EWL states belong to
the class of the `X' states. Explicitly, if the density matrix of a quantum
state is of the form
\begin{eqnarray}
\left(
\begin{array}{cccc}
\rho _{11} & 0 & 0 & \rho _{14} \\
0 & \rho _{22} & \rho _{23} & 0 \\
0 & \rho _{23}^{\ast } & \rho _{33} & 0 \\
\rho _{14}^{\ast } & 0 & 0 & \rho _{44} \\
&  &  &
\end{array}%
\right) ,
\end{eqnarray}
then it belongs to the class of the X states.

The EWL states have the following advantages. First, we can easily find that
if the initial state of two atoms is X structure, then the reduced density
matrix at arbitrary time t is still X structure under the single atom
evolution determined by the master equation (4) (in the basis $|1\rangle
=|++\rangle ,|2\rangle =|+-\rangle ,|3\rangle =|-+\rangle ,|4\rangle
=|--\rangle $). Second, the EWL states allow us to clearly show the
influence of the purity and the amount of entanglement of the initial states
on the entanglement dynamics of two atoms simultaneously. The purity of the
EWL states are dependent on the parameter $p$ and the amount of the
entanglement of the EWL states are related to $\mu $ and $\nu $. If $p=1$
the EWL states reduce to the Bell-like states $|\Phi \rangle $ and $|\Psi
\rangle $. In the case of $p=1,\mu =\nu =1/\sqrt{2}$ the EWL states become
the Bell states while in the case of $p=0$ they are the maximally mixed
states. Third, as we will see, it is easy for us to calculate the
entanglement dynamics of two atoms which are initially prepared in EWL
states.

\section{Entanglement and coherence}
In this section, we will analyze the entanglement dynamics and coherence of
two atoms by employing the concurrence and linear entropy, respectively.
In order to study the entanglement of above system described by density
matrix $\rho $, we adopt the measure concurrence which is defined by \cite%
{Wootters1998}
\begin{equation}
C=\max {\{0,\lambda _{1}-\lambda _{2}-\lambda _{3}-\lambda _{4}\}},
\end{equation}%
where the $\lambda _{i}$ ($i=1,2,3,4$) are the square roots of the
eigenvalues in decreasing order of the magnitude of the \textquotedblleft
spin-flipped" density matrix operator $R=\rho (\sigma _{y}\otimes \sigma
_{y})\rho ^{\ast }(\sigma _{y}\otimes \sigma _{y})$ and $\sigma _{y}$ is the
Pauli Y matrix, i.e., $\sigma _{y}=\left(
\begin{array}{cc}
0 & -i \\
i & 0%
\end{array}%
\right) $. Particularly, for X structure states defined by Eq. (17), it is
easy to find the concurrence is
\begin{widetext}
\begin{eqnarray}
C(t)=2\max{\{0,|\rho_{23}(t)|-\sqrt{\rho_{11}(t)\rho_{44}(t)},
|\rho_{14}(t)|-\sqrt{\rho_{22}(t)\rho_{33}(t)}\}}\label{c1}
\end{eqnarray}
\end{widetext}Combing Eqs. (15), (16), and (19), we find the concurrence of
the two atoms, which are initially prepared in EWL states ($\rho _{\Phi }$
or $\rho _{\Psi }$), are
\begin{eqnarray}
C_{\Phi }(t) &=&2\max \{0,|f(t)\lambda (t)|^{2}|\rho _{23}(0)|-\sqrt{\rho
_{11}(0)\rho _{44}(0)}\},  \nonumber \\
C_{\Psi }(t) &=&2\max \{0,|f(t)\lambda (t)|^{2}|\rho _{14}(0)|-\sqrt{\rho
_{22}(0)\rho _{33}(0)}\}.  \nonumber \\
&&
\end{eqnarray}%
Inserting the initial state of two atoms of Eq.(16) into the above equations leads
to
\begin{eqnarray}
C(t) &=&C_{\Phi }(t)=C_{\Psi }(t)  \nonumber  \label{concurrence} \\
&=&\max {\{0,2p|f(t)\lambda (t)|^{2}|\mu \nu |-\frac{1-p}{2}\}}.
\end{eqnarray}%
We find the concurrence $C_{\Phi }(t)$ and $C_{\Psi }(t)$ are the same if
the parameter $\mu $ and $\nu $ of $\rho _{\Phi }$ and $\rho _{\Psi }$ are
equal, respectively. It is clear that the entanglement of the two atoms depends on the
initial state of the them, i.e., $C(t)$ relies on the parameters $p$, $\mu $%
, and $\nu $.

In Fig. 2 and Fig. 3, the concurrence $C(t)$ is  plotted as a function of
the dimensionless scaled time $\Omega t$ and the parameter $k/\Omega $ for
different values of $p$. From these two figures, one can easily find that
the entanglement of the two atoms suddenly goes to zero and stays zero for a
finite time, i.e., the ESD phenomenon appears in the present system.
However, the entanglement cannot revive completely due to the dissipation of
the cavity fields. The lower panels of Fig.2 and Fig.3 are the contour plots
of the concurrence, where the severe shading areas indicating the two atoms
are completely disentangled. Comparing the upper panels of Fig.2 and Fig.3,
one can clearly see that the entanglement dynamics relies heavily on the
purity of the initial state which is represented by the parameter $p$. For
instance, the maximal value of concurrence in Fig.2 is about $0.7$ while the
maximal value of concurrence in Fig.3 is only about $0.4$. The contour plots
of these figures clearly show that the areas showing entanglement
significantly decrease with the decrease of the purity of the initial state.
To show this more intuitively, we plot the concurrence as a function of the
dimensionless scaled time $\Omega t$ for several values of $p$ in Fig.4.
Also, the entanglement dynamics depends on the initial state of the field $%
|\alpha \rangle $ as one can see from the upper panel of Fig.5. The
concurrence decreases with the increase of the amplitude of $\alpha $. For
instance, in the case of $k/\Omega =0.01$, $\mu =\nu =1/\sqrt{2}$, $p=0.9$,
and $\alpha =0.5$, there is no ESD. However, in the case of $\alpha =1$ or $%
\alpha =2$, the ESD phenomenon could appear.

In order to see the time limits (long-term) behavior of the entanglement we
plot the concurrence as a function of the dimensionless scaled time $\Omega t
$ ($0\leqslant \Omega t\leqslant 500$) for $\alpha =0.5$ (solid line), $\alpha =1$
(dashed line), and $\alpha =2$ (dotted line) in the lower panel of Fig.5.
From this figure, we see that the entanglement of the two atoms survives for
a long time if the mean photon number of the fields $|\alpha |^{2}$ has
small values (see the solid and dashed lines of Fig.5). The entanglement of
the stationary state $\rho (\infty )$ depends on the initial state of the
two atoms and the ratio of the decay rate of two cavities to the atom-field
coupling constant. If the initial state is completely mixed ($p=0$), that
is, the two atoms are disentangled initially, then no entanglement will be
generated in the system.

In Fig.6, we have presented the time-evolution of entanglement as a
function of $kt$ and $\Omega /k$. One can see clearly that the entanglement
decreases gradually as $kt$ increases for a given time. In principle, by
using the explicit expression of the concurrence, one can find exactly the
threshold value after which a long-lived entanglement can be obtained. It is
easy to see that the oscillations of the concurrence gradually disappears as
the time $kt$ increases. Actually, from Eq.(21) one can see that $C(t)=c$ ($%
c$ is a constant) leading to the following condition: $kt>kt_{c}$ which can
be obtained from the following equation(we set $\mu=\nu=1/\sqrt{2}$)
\begin{equation}
{|f(t)|}=\frac{{1}}{|\langle \alpha _{-}(t)|\alpha _{+}(t)\rangle |}\sqrt{\frac{%
2c+1-p}{2p}},
\end{equation}%
then, the entanglement has no chance to be oscillating when $kt$ exceeds a
threshold value $kt_{c}$ (from Fig. 6, $kt_{c}\approx7)$. Here we should point out
that the effective threshold value of the decay time corresponding to the
transition of the entanglement from oscillating regime to fully constant
value is independent of the interaction time.
 From the above discussions, one can observe that
the initial entanglement of the two atoms could be partially preserved in
the present system which is useful for quantum information processing and
quantum memory.

In order to show the dependence of
the long time behavior of the entanglement of the atoms  on the parameter
$\alpha$ and the purity of the initial state $p$, we plot the concurrence
as a function of $\alpha$ and $p$ with $t=500$ in Fig.7. From Fig.7, we observe that
the stationary state entanglement of the atoms decreases with the increase of the
parameter $\alpha$ while increases with the purity of the initial state $p$.
This observation is consistent with the previous discussions.
We have also carefully investigated the entanglement
dynamics of the system if only one cavity is subjected to decay radiation
field and find that the entanglement will also be partially preserved.

Now, we investigate the coherence of the atoms by employing the linear
entropy defined by \cite{Zurek1993,Lidar2003}
\begin{equation}
S(t)=1-Tr[\rho ^{2}(t)].
\end{equation}%
The linear entropy $S$ is zero for pure states and 1 for completely mixed
states. Tracing out the degrees of the freedom of the second atom from Eq.
(17) and using Eq. (15), we get the reduced density matrix of atom 1
\begin{eqnarray}
\rho _{1}(t) &=&[\rho _{11}(0)+\rho _{22}(0)]|e\rangle \langle e|  \nonumber
\\
&&+[\rho _{33}(0)+\rho _{44}(0)]|g\rangle \langle g|.
\end{eqnarray}%
Thus the linear entropy of atom 1 is
\begin{equation}
S_{1}(t)=S_{1}(0)=1-Tr[\rho _{1}^{2}(t)]=2|\mu \nu |^{2}>0.
\end{equation}
Similarly, for atom 2, we have
\begin{equation}
S_{2}(t)=S_{2}(0)=1-Tr[\rho _{2}^{2}(t)]=2|\mu \nu |^{2}>0.
\end{equation}
From the above equations, we find that the coherence of the atoms is
preserved in the present model, that is, the subspace of the atoms are
decoherence-free subspaces \cite{Lidar2003}.

\section{Conclusions}
In the present paper, we have investigated the entanglement dynamics and
coherence of a quantum system formed by two two-level atoms interacting with
two spatially separated and dissipative cavities in the dispersive limit.
With the help of superoperator method, we obtained an explicit expression of
the reduced density matrix of the two atoms and calculated the entanglement
and coherence of them by employing the concurrence and linear entropy,
respectively. In a short-term regime, the ESD phenomenon could be obtained and
the entanglement of the two atoms decreases with the time development.
However, in the long-term, the entanglement of the two atoms tends to a
fixed value. This value depends only on the initial state of the two atoms,
the atom-field coupling constant, and the decay rate of fields.
Particularly, we found that there is long-lived entanglement (or stationary
state entanglement) in the presence of the dissipation of the fields. In
other words, in the dispersive limit, the initial entanglement of the two
atoms within two dissipative cavities could be partially preserved for a
long time.

Finally, we calculated the coherence of the two atoms using the linear
entropy. Our results show that the coherence of each atom can be preserved.
Thus the subspace of each atom is a decoherence-free subspace. In a word,
the initial coherence (entanglement) of the two atoms could be preserved
(partially preserved) in the dispersive limit even if they are put into two
dissipative cavities. This feature is useful for quantum information
processing and quantum memory.
It is interesting to extend the present system to a many-qubit system.
Note that Eq.(15) is not suitable for the many-qubit system. However,
the method introduced by Bellomo, France, and Compagno \cite{Bellomo2008}
is still applicable. Thus, one can investigate whether the multipartite
entanglement of the many-qubit system can be preserved.

\section*{Acknowledgments}
 This project was supported by the Natural Science
Foundation of Jiangxi, China (Grant No 2007GZW0819 and 2008GQW0017),
and the Scientific Research Foundation of Jiangxi Provincial
Department of Education (Grant No GJJ09504), and the Foundation of
Talent of Jinggang of Jiangxi Province (Grant No 2008DQ00400).

\newpage
\begin{figure}[tbp]
\centering {\scalebox{0.7}[0.7]{\includegraphics{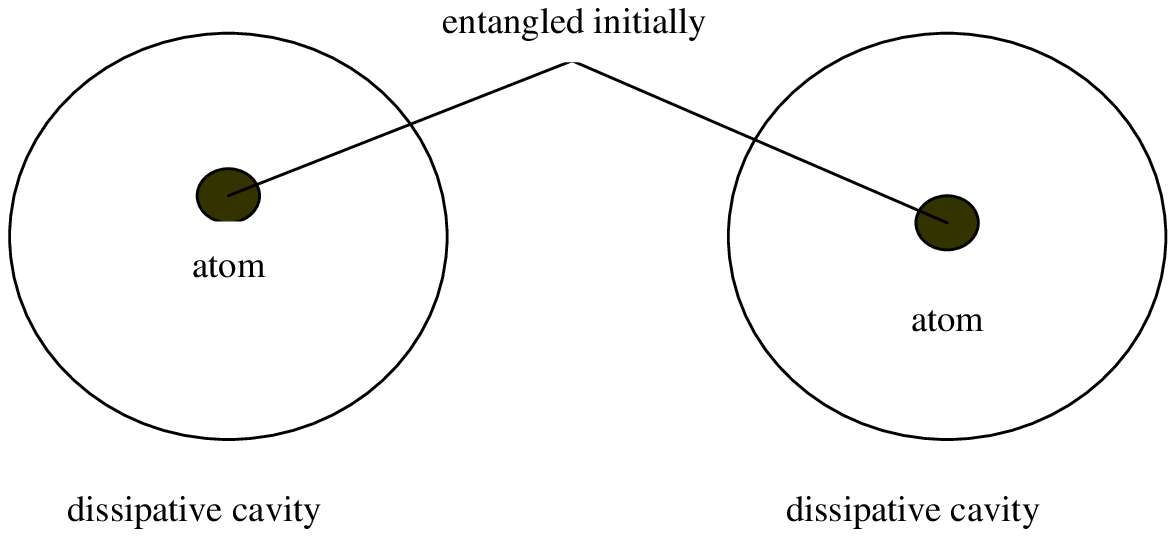}}}
\caption{ The schematic picture of two two-level atom within two spatially
separated and dissipative cavities. The two atoms are initially prepared in
EWL states. Each atom is put into a single-mode cavity and interacts with
its own cavity field locally. Note that there is no direct interactions
between two atoms once they have been put into cavities. }
\end{figure}

\begin{figure}[tbp]
\centering {\scalebox{1.5}[1.5]{\includegraphics{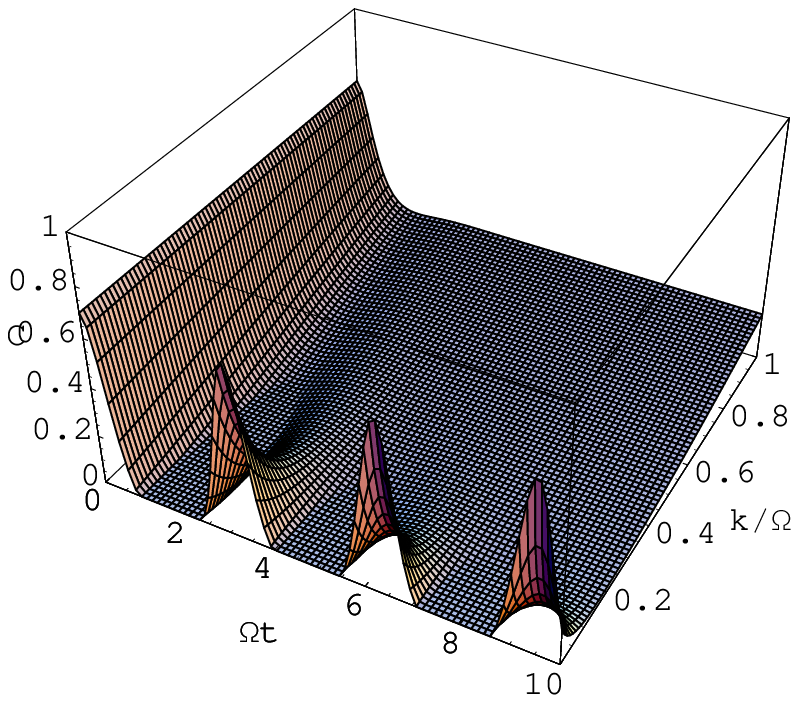}}}
\caption{The concurrence of two atoms is plotted as a
function of the dimensionless scaled time $\Omega t$ and parameter $%
k/\Omega$ with $\protect\alpha=1$, $\protect\mu=\protect\nu=1/\protect\sqrt{2%
}$, and $p=0.8$.}
\end{figure}

\begin{figure}[tbp]
\centering {\scalebox{1.5}[1.5]{\includegraphics{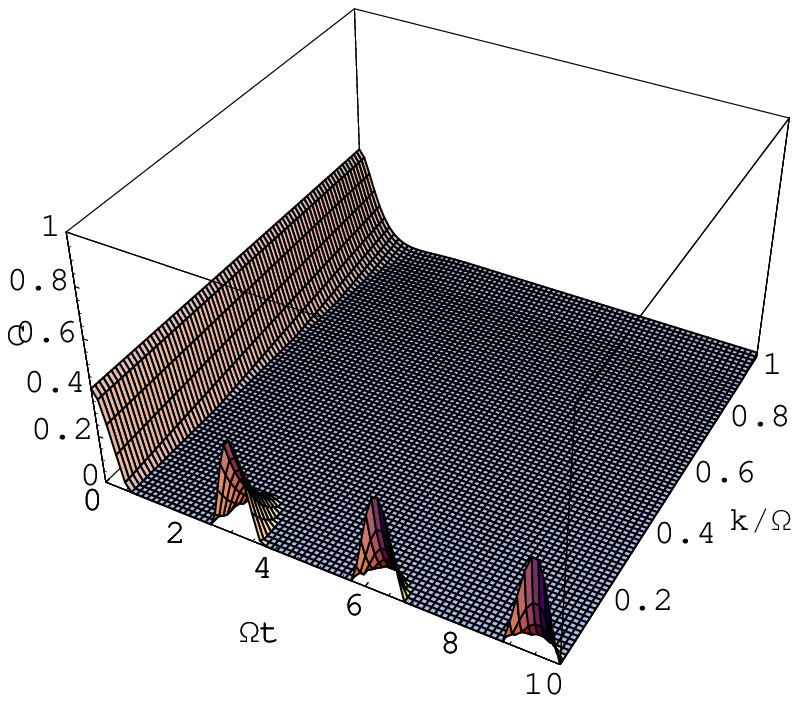}}}
\caption{The concurrence of two atoms is plotted as a
function of the dimensionless scaled time $\Omega t$ and parameter $%
k/\Omega$ with $\protect\alpha=1$, $\protect\mu=\protect\nu=1/\protect\sqrt{2%
}$, and $p=0.6$.  }
\end{figure}

\begin{figure}[tbp]
\centering {\scalebox{1.2}[1.5]{\includegraphics{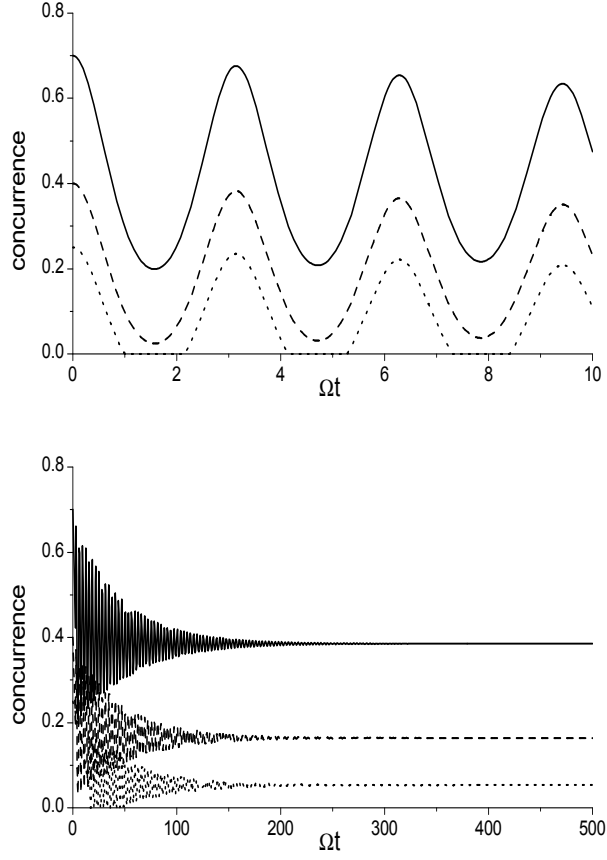}}}
\caption{The concurrence of two
atoms is plotted as a function of the dimensionless scaled time $\Omega t$
with $k/\Omega=0.01$, $\protect\alpha=0.5$, $\protect\mu=\protect\nu=1/\protect%
\sqrt{2}$ for $p=0.8$(solid line), $p=0.6$(dashed line), and $p=0.5$(dotted
line).
Upper panel: The short-term behavior of the entanglement between two atoms.
Lower panel: The long-term behavior of the entanglement between two atoms.}
\end{figure}

\begin{figure}[tbp]
\centering {\scalebox{1.2}[1.5]{\includegraphics{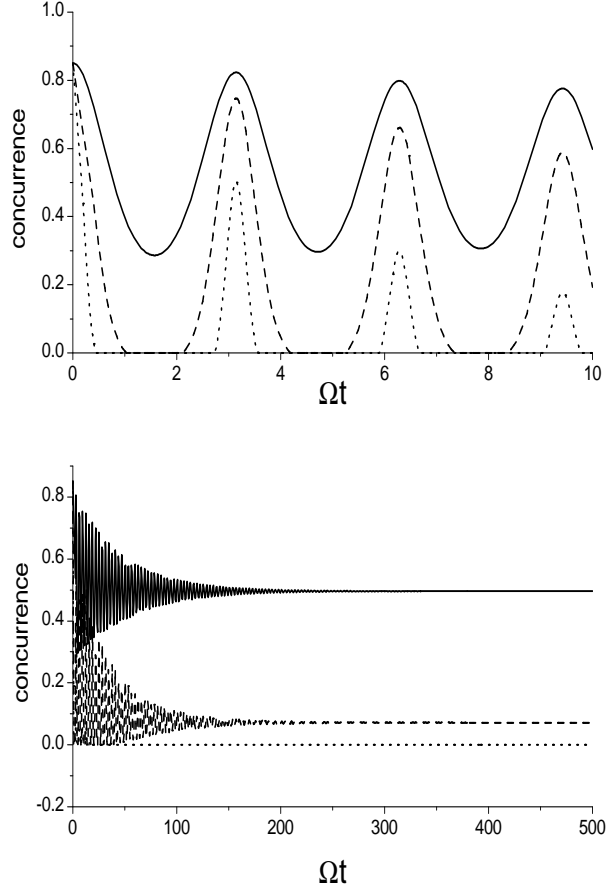}}}
\caption{ The concurrence of two atoms is plotted as a function of the
dimensionless scaled time $\Omega t$ with $k/\Omega=0.01$, $p=0.9$, $\protect%
\mu=\protect\nu=1/\protect\sqrt{2}$ for $\protect\alpha=0.5$(solid line), $%
\protect\alpha=1$(dashed line), and $\protect\alpha=2$(dotted line). Upper
panel: The short-term behavior of the entanglement between two atoms. Lower
panel: The long-term behavior of the entanglement between two atoms.
Comparing the upper panel with the lower panel, one can observe that, in the
short-term, there is ESD and the maximal values of entanglement decrease
with time. However, in the long-term, there is long-lived entanglement(see
the solid and dashed lines of the lower panel). The entanglement decreases
with the increase of the mean photon number of the field represented by $|%
\protect\alpha|^2$. If the parameter $|\protect\alpha|^2$ is larger enough,
there is no long-lived entanglement(see the dotted line of the lower panel).}
\end{figure}

\begin{figure}[tbp]
\centering {\scalebox{1.5}[1.5]{\includegraphics{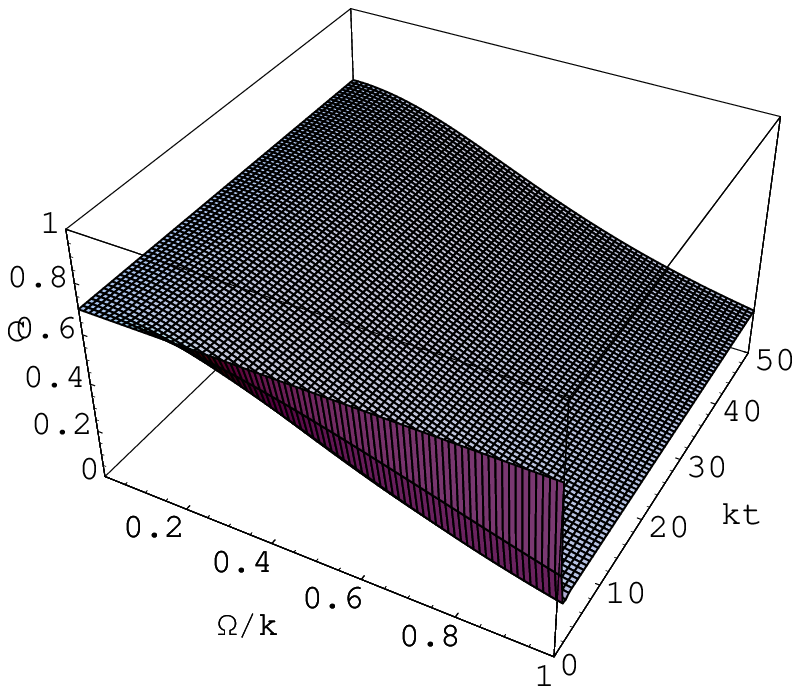}}}
\caption{ The concurrence of two
atoms is plotted as a function of the dimensionless scaled time $k t$
and parameter $\Omega/k$ with $\protect\alpha=1$, $\protect\mu=\protect\nu=1/\protect%
\sqrt{2}$ for $p=0.8$. }
\end{figure}

\begin{figure}[tbp]
\centering {\scalebox{1.5}[1.5]{\includegraphics{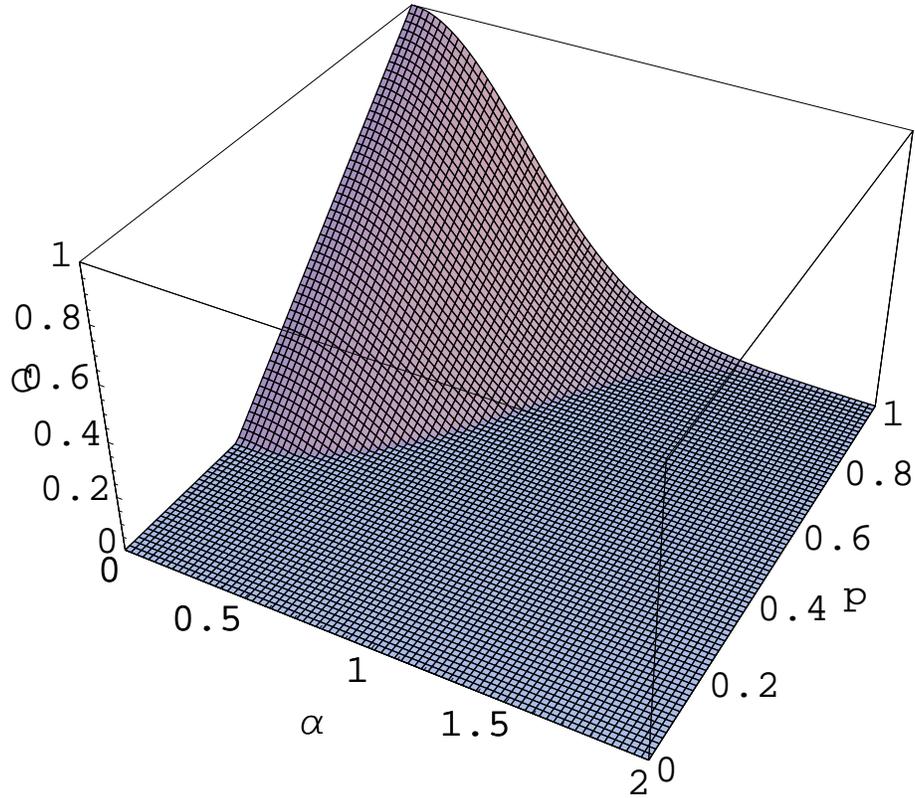}}}
\caption{ The concurrence of two
atoms is plotted as a function of the parameter $\alpha$
and  $p$ with $\Omega=1,k=0.01$, $\protect\mu=\protect\nu=1/\protect%
\sqrt{2}$, and $t=500$. This figure clearly shows the dependence of
the long time behavior of the entanglement of the atoms on the parameter
$\alpha$ and the purity of the initial state $p$.
From this figure, we see that the stationary state entanglement of the atoms decreases with
the parameter $\alpha$ while increases with the purity of the initial state $p$.}
\end{figure}

\end{document}